\documentclass[letterpaper, 10 pt, conference]{ieeeconf}  

\IEEEoverridecommandlockouts                              
\usepackage{graphicx} 
\usepackage{epsfig} %
\usepackage{times} 
\usepackage{amsmath}
\usepackage{amssymb}  
\usepackage{svg}
\usepackage{algorithm}
\usepackage{algpseudocode}
\usepackage{booktabs} 
\usepackage{stfloats}
\usepackage{float} 
\usepackage{url}
\usepackage{fixltx2e}
\usepackage{svg}
\usepackage{tabularx}
\usepackage{fancyhdr}
\usepackage{hyperref}
\usepackage{placeins}

\newtheorem{remark}{\bf Remark}

\algdef{SE}[SUBALG]{Indent}{EndIndent}{}{\algorithmicend\ }%
\algtext*{Indent}
\algtext*{EndIndent}

\newcommand{\Lone}{$\mathbf{L}_{1}$}
\newcommand{\Ltwo}{$\mathbf{L}_{2}$}
\newcommand{\Linf}{$\mathbf{L}_{\infty}$}

\newcommand{\IV}{\ensuremath{IV}}

\title{\LARGE \bf
ANOCA: AC Network-aware Optimal Curtailment Approach for Dynamic Hosting Capacity
}
\author{Emmanuel O. Badmus, and Amritanshu Pandey
\thanks{This material is based upon work supported by the U.S. Department of Energy’s award number DE-EE0010407 and award number DE-AC02-05CH11231. The views expressed herein do not necessarily represent the views of the U.S. Department of Energy or the United States Government.}
\thanks{E. Badmus and A. Pandey are both with the Department of Electrical and Biomedical Engineering at the University of Vermont, Burlington, VT. {\tt\small {Ebadmus, Apandey1}@auvm.edu}}%
}
\begin{document}
\maketitle

\thispagestyle{fancy}
\pagestyle{fancy}
\fancyhf{} 
\renewcommand{\headrulewidth}{0pt}
\renewcommand{\footrulewidth}{0pt}
\fancyfoot[C]{PREPRINT Version: Prepared for the 63rd Conference on Decision and Control.}

\begin{abstract}

With exponential growth in distributed energy resources (DERs) coupled with at-capacity distribution grid infrastructure, prosumers cannot always export all extra power to the grid without violating technical limits. Consequently, a slew of dynamic hosting capacity (DHC) algorithms have emerged for optimal utilization of grid infrastructure while maximizing export from DERs. Most of these DHC algorithms utilize the concept of \textit{operating envelopes (OE)}, where the utility gives prosumers technical power export limits, and they are free to export power within these limits. Recent studies have shown that OE-based frameworks have drawbacks, as most develop power export limits based on convex or linear grid models. As OEs must capture extreme operating conditions, both convex and linear models can violate technical limits in practice because they approximate grid physics. However, AC models are unsuitable because they may not be feasible within the whole region of OE. We propose a new two-stage optimization framework for DHC built on three-phase AC models to address the current gaps. In this approach, the prosumers first run a receding horizon multi-period optimization to identify optimal export power setpoints to communicate with the utility. The utility then performs an infeasibility-based optimization to either accept the prosumer's request or dispatch an optimal curtail signal such that overall system technical constraints are not violated. To explore various curtailment strategies, we develop an \Lone, \Ltwo, and \Linf norm-based dispatch algorithm with an exact three-phase AC model. We test our framework on a 1420 three-phase node meshed distribution network and show that the proposed algorithm optimally curtails DERs while guaranteeing the AC feasibility of the network.

\end{abstract}

\section*{Nomenclature}
\noindent
\begin{tabularx}{\textwidth}{@{}>{\hsize=0.2\hsize}p{8cm}X@{}}
\textbf{Symbol} & \textbf{Interpretation} \\
$P_{i,\tau}, P_{e,\tau}$ & Power imported from/exported to the grid \\
$P_{c,\tau}, P_{d,\tau}$ & Battery charging/discharging power \\
$E_{b,\tau}$ & State of charge (SOC) of the battery \\
$P_{l,\tau}, Q_{l,\tau}$ & Real and reactive power demand of the load \\
$P_{pv,\tau}$ & Power generated by the PV system \\
$C_{i,\tau}, C_{e,\tau}$ & Cost of power import/export to/from grid \\
$E_{\text{set}}$ & SOC setpoint of the battery at start/end \\
$\eta_{c}, \eta_{d}$ & Battery efficiencies (charging, discharging) \\
$\overline{P}, \overline{E}$ & Max power, SOC limits \\
$\mathcal{T}, \Delta \tau$ & Optimization horizon and time step \\
$V^{R}, V^{I}$ & Real and Imag. parts of the voltage, $\in \mathbb{R}^{n}$ \\
$I^{R}, I^{I}$ & Real and Imag. parts of the current, $\in \mathbb{R}^{n}$ \\
\end{tabularx}

\noindent
\begin{tabularx}{\textwidth}{@{}>{\hsize=0.2\hsize}p{8.0cm}X@{}}
$\mathcal{N}, \mathcal{N}_l, \mathcal{N}_{pv}$ & Set of all nodes, a subset of load nodes, \newline and a subset of nodes with PV generation \\
$\Phi, \phi, \phi'$ & Set of phases, active phase, mutual phases \\
$G_{nm,\phi,\phi'}$ & Conductance between $n$ and $m$ for $\phi$ and $\phi'$ \\
$B_{nm,\phi,\phi'}$ & Susceptance between $n$ and $m$ for $\phi$ and $\phi'$ \\
$V^{R}_{n,\phi}, V^{I}_{n,\phi}$ & Real and Imag. voltage at node $n$ for phase $\phi$\\
$V^{R}_{nm,\phi}, V^{I}_{nm,\phi}$ & Real and Imag. parts of the voltage difference \newline between nodes $n$ and $m$ on phase $\phi$ \\
$I^{R}_{nm,\phi}, I^{I}_{nm,\phi}$ & Real and Imag. parts of the current flow on \newline line $l$ between nodes $m$ and $n$ on phase $\phi$ \\
$P_{n,\phi}$ & Power injection at node $n$ for phase $\phi$ \\
$P_{i,n,\phi}, P_{e,n,\phi}$ & Import/export power at node $n$ for phase $\phi$ \\
$P_{l,n,\phi}, Q_{l,n,\phi}$ & Real/reactive load at node $n$ for phase $\phi$ \\
$P_{cu,n,\phi}$ & Curtailed power at node $n$ for phase $\phi$ \\
$\overline{I}_{l,\phi}$ & Max. allowable current for line for phase $\phi$ \\
$t_{k}, \underline{t}_{k}, \overline{t}_{k}$ & Tap ratio and min/max tap limits for Xfmr $k$ \\
$\underline{V}_{k}, \overline{V}_{k}$ & Min/max voltage limits for Xfmr $k$ \\
$\underline{V}_{n,\phi}, \overline{V}_{n,\phi}$ & Min/max voltage limits at node $n$ for phase $\phi$ \\
$P_{k,fl}, Q_{k,fl}$ & Real and reactive power on Xfmr $k$, line $fl$ \\
$\overline{S}_{k,fl}$ & Max apparent power of Xfmr $k$ on line $fl$ \\
\end{tabularx}

\section{INTRODUCTION}

\vspace{-0.5em}
\subsection{Research Motivation}
\vspace{-0.2em}
\noindent Historically, electric power flow was predominantly unidirectional, from mostly large centralized generation to the distribution grid consumers through an extensive network of high-voltage and low-voltage electric networks. In this paradigm, the grid operators positioned consumers as passive participants, solely responsible for paying for consumed electricity without actively participating in exporting any extra power to the grid. However, this traditional paradigm is becoming inoperative due to the proliferation of distributed energy resources (DERs), especially rooftop photovoltaics (PVs), with many consumers feeding power back into the grid. As such, in many distribution feeders across the globe, we are reaching a tipping point wherein, without active control, any marginal increase in net export of power will result in technical violation of the grid limits.
\vspace{-0.2em}




\subsection{State-of-the-art Solutions and Limitations}
\vspace{-0.2em}
\noindent One of the first proposed solutions by utilities to maintain grid operation within technical limits (with large penetration of rooftop solar) is termed static hosting capacity (SHC) \cite{bollen2005power}, which offers a firm connection to the customers with no risk of curtailment. 
However, once a \textit{static-limit} is hit, new PV systems cannot export to the grid unless they pay for system upgrades. The static limit is based on \textit{select worst-case operating conditions} and, consequently, very conservative \cite{capitanescu2014assessing}, resulting in a significant loss of PV production.

\begin{figure}[htpb]
    \centering
    \includegraphics[width=0.93\linewidth]{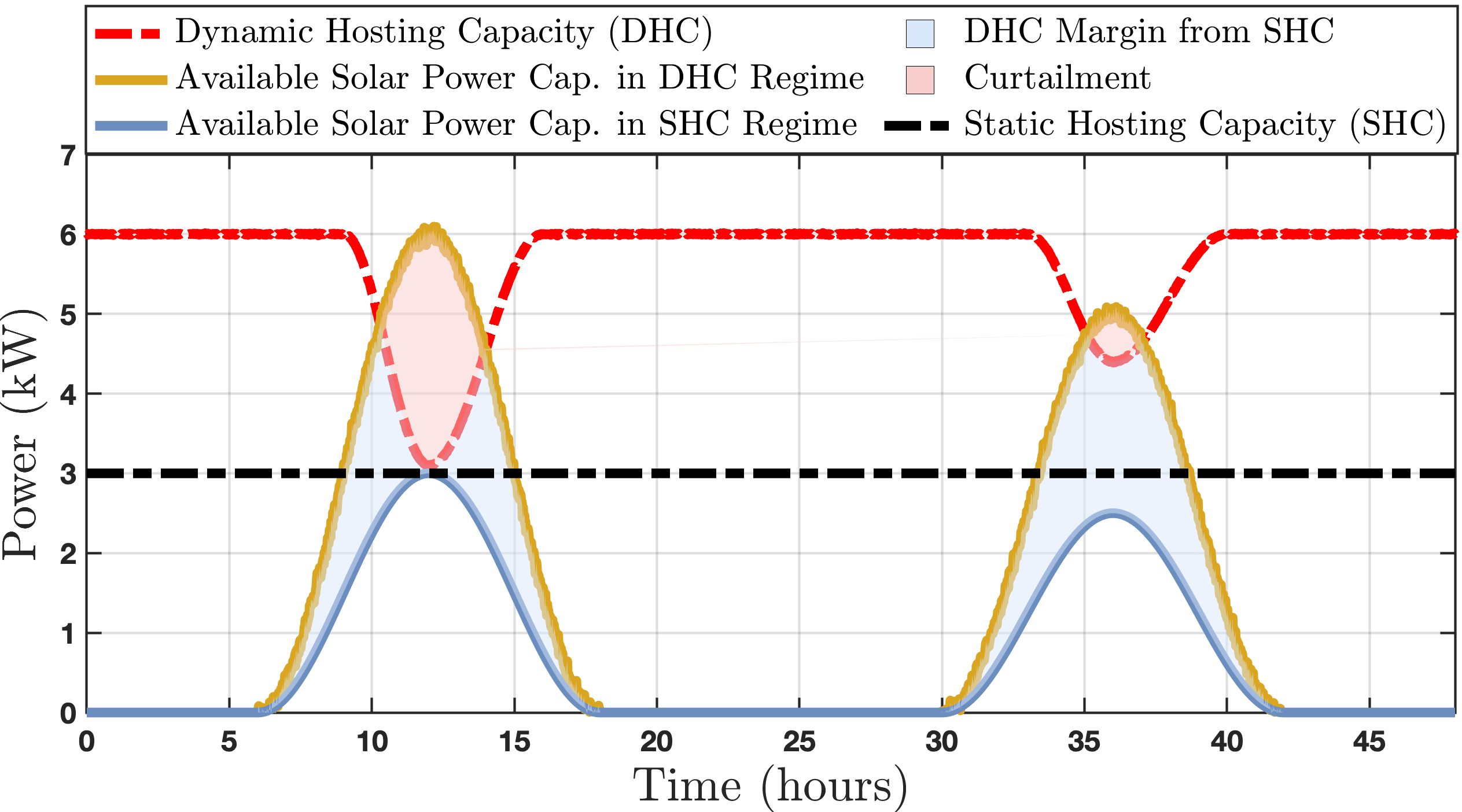}
    \caption{Comparison of Dynamic Hosting Capacity (DHC) and Static Hosting Capacity (SHC) paradigms. The dashed black line represents the SHC limit and the available PV export capacity in blue. The dashed red line represents the DHC limit, and the available PV export capacity is in yellow. Note the PV installation sizes in SHC vs. DHC scenarios are different.}
    \label{fig:DHC_SHC}
\end{figure}

To counter the inefficiencies of SHC and allow more PVs to participate actively, algorithms under the broad umbrella term of dynamic hosting capacity (DHC) have been proposed. This technology is network-aware and dynamic, assuming i) access to distribution network models and ii) information flow between consumers and the utility  (with communication standards such as IEEE 2030.5). 
\noindent The fundamental idea of DHC is that utilities will propose time-varying operating limits, which the various PV-exporting consumers must adhere to \cite{liu2022using, yi2022fair, astero2018improving, Liu_2023, gebbran2022multiperiod, mahmoodi2021hosting}. 
These utilities calculate optimal export (OE) limits for active consumers for predetermined time intervals, integrating controllable network components, such as batteries and transformer taps. Fig.\ref{fig:DHC_SHC} illustrates the two paradigms of hosting capacity, underscoring the utility's role in managing network constraints and facilitating renewable integration.

\vspace{0.3em}
\vspace{4mm} 

\vspace{-0.2em}
The DHC OE values in most current frameworks  \cite{liu2022using, yi2022fair, astero2018improving, Liu_2023, gebbran2022multiperiod, mahmoodi2021hosting} are based on linear and convex relaxation models, which do not satisfy AC network constraints in most practical scenarios. 
Subsequently, this discrepancy in DHC OE values between AC versus relaxed and linear models is significant at the boundary operating conditions, which may violate voltage and flow limits during DHC enforcement in practice \cite{moring2023inexactness}. 
New approaches (e.g., see \cite{nazir2021grid} that uses convex restriction) address the underlying challenge; however, they underutilize the grid capacity and may not apply to meshed networks without loss of generality. 
\vspace{-0.4em}
\subsection{Re-envisioning Dynamic Hosting Capacity with ANOCA}
\begin{figure*}[htpb]
    \centering
    \includegraphics[width=0.90\linewidth]{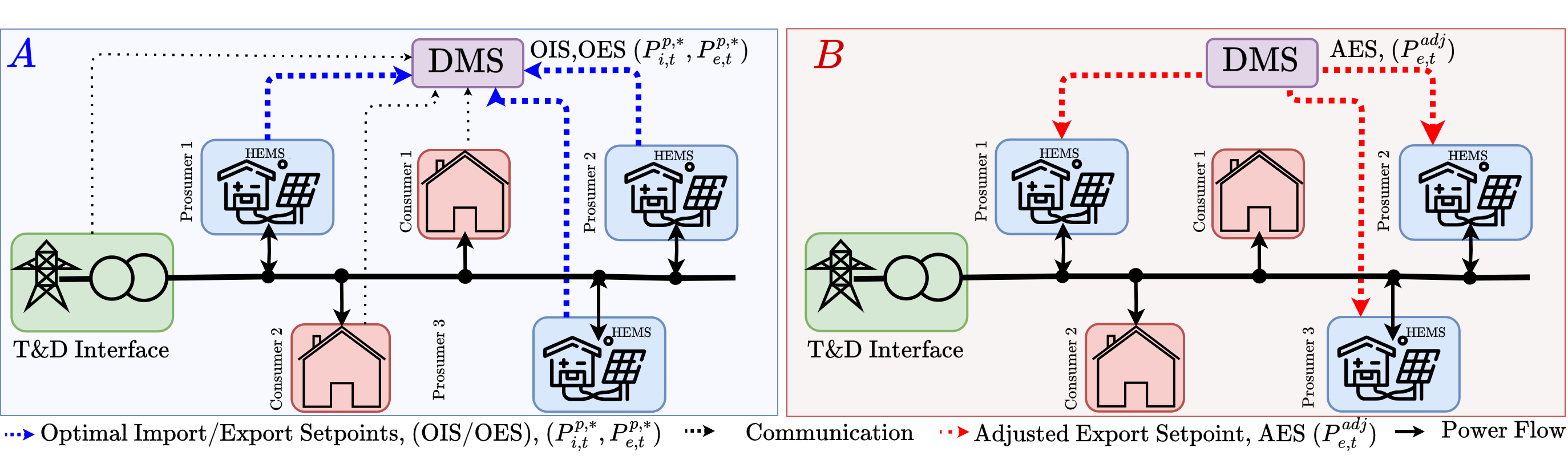}
    \caption{Two-stage schematic of the ANOCA framework. First, in stage A, prosumers' HEMS will calculate their optimal export setpoints (OES) for time-step $t$ and broadcast it to the utility DMS (see left of the figure). Next, in stage B, the DMS runs an infeasibility-based optimization to accept prosumers' OES or enforce adjusted export setpoints (AES) using curtailments (see right of the figure).}
    \label{fig:proposed_framework}
\end{figure*}

\vspace{-0.3em}
\noindent Learning from current challenges in the DHC paradigm, we assert that using three-phase AC unbalanced network models for DHC can have significant benefits. 
We propose a new two-stage DHC paradigm for that goal, the AC network-aware optimal curtailment approach-ANOCA, illustrated in Fig \ref{fig:proposed_framework}. 
In the ANOCA paradigm, instead of the utility providing prosumers with time-varying OEs, prosumers first calculate \textit{optimal export setpoints (OES)} with home energy management systems (HEMS) and broadcast them to the utility. 
Next, the utility processes the proposed optimal export setpoints in the advanced distribution management system (DMS) and issues either a go-ahead signal (no curtailment necessary) or an \textit{adjusted export setpoint (AES)} (with curtailment). 
We design the proposed two-stage framework to ensure i) efficiency (i.e., optimal utilization of network capacity) and feasibility (i.e., making curtailment decisions based on AC constraints). 
In our approach, we formulate the HEMS problem as a receding horizon multiperiod optimization, and we extend a grid infeasibility-focussed idea \cite{foster2022three} to formulate the DMS problem.

To develop and explore equitable and sparse curtailment strategies in ANOCA, we formulate the DMS optimization with differing objectives, minimizing the \Lone, \Ltwo, and \Linf{} norm of the magnitudes PV curtailment. 
The paper re-envisions and contributes to the literature by developing a DHC algorithm that has the following features:
\vspace{-0.2em}
\begin{enumerate}[]
    \item \textbf{AC feasibility}: Ensure the PV export always satisfies AC network constraints without technical violations.
    \item \textbf{Equitable curtailment:} \Linf-norm objective in the DMS minimization results in the most equitable curtailment (maximum curtailment by any given prosumer is lowest) but incurs a high volume of net curtailment ($\sim$ 20x of curtailments in \Lone{} and \Ltwo{} strategy for a meshed network).  
    \item \textbf{Sparse curtailment:} \Lone-norm objective in the DMS minimization results in the most sparse curtailment (few prosumers curtail a large amount).  
    \item \textbf{Balanced curtailment}: \Ltwo-norm objective in the DMS minimization promotes a balanced distribution of curtailments, achieving a trade-off between the sparse but high magnitude curtailments and equitable but large net curtailments.
\end{enumerate}
\vspace{-0.5em}

\section{PRELIMINARIES}
\subsection{Three-phase Distribution Grid Optimization}
\vspace{-0.2em}
\noindent In the proposed ANOCA framework, we use the \IV{} based three-phase equivalent circuit model of the grid \cite{jereminov2016equivalent}. The model is equivalent to the current-injection model \cite{garcia2000three} in its native form. Given an undirected graph representation of the distribution grid, $\mathcal{G}$($\mathcal{N}$, $\mathcal{E}$), the \IV{} model represents the relationship between various edge $\mathcal{E}$ and node $\mathcal{N}$ features using Kirchhoff's current and voltage laws. It uses the real and imaginary parts of rectangular coordinates to describe the grid's complex voltages $(V_{\phi}^R, V_{\phi}^I)$ and current flows $(I_{\phi, fl}^R, I_{\phi,fl}^I)$, where $\phi$ is from the set of three-phases ${a, b, c}$. Importantly, the network constraints from edge elements, including transmission lines, cables, and transformers, are linear. Nonlinearities only stem from power injections and other controls. Note that the \IV{} based formulation captures the AC physics of the network components exactly and applies to both radial and meshed networks without loss of generality. Current works use this model for grid optimizations \cite{foster2022three},\cite{jereminov2018equivalent}. Mathematically, the problem is non-convex; however, it has been shown to converge to \textit{good} feasible solutions consistently. 
In Section \ref{subsec:math-form-DMS}, we will use this model to formulate the DMS problem, including the different curtailment strategies.

\subsection{HEMS-based Dynamic Hosting Capacity}
\vspace{-0.5em}
\noindent 
There are prior works on a two-stage HEMS-based strategy for DHC algorithms (most relevant include \cite{yi2022fair}, \cite{iria2022mv}, \cite{petrou2021ensuring}). \cite{yi2022fair} incorporates a two-stage optimization process to enhance grid integration of DERs. 
In the initial stage, the utility uses second-order conic grid models to develop OEs for each prosumer. 
In the second stage, each prosumer optimizes their power export and import with HEMS OEs from the utility.
More relevant to this work, \cite{iria2022mv}  details a two-stage algorithm where the aggregator optimizes bidding strategies for real-time energy and reserve markets, ensuring the security of both MV and LV networks. 
Also relevant, \cite{petrou2021ensuring} outlines dynamic operating limits for prosumers to maintain distribution network integrity, utilizing a two-stage framework that dynamically adjusts prosumer exports based on real-time network conditions. These works inspire the high-level structure of the proposed work, but our methodology diverges significantly regarding the grid models employed and the specific optimization strategies implemented.

\section{Problem Formulation}

\vspace{-0.2em}
\noindent We propose the ANOCA two-stage optimization framework (see Fig. \ref{fig:proposed_framework}) to maximize PV export without violating AC grid constraints and overcome challenges with current DHC approaches. 
Our approach requires minimal information exchange between prosumers and the central DMS operator and uses a full three-phase AC model to make curtailment decisions. 
The recursive DHC algorithm begins with individual prosumers calculating an optimal export setpoint (OES) for each time interval $\Delta t$ from a receding horizon multi-period optimization performed by HEMS.
Subsequently, the OESs from various prosumers for the current time step $t$ are broadcasted to the DMS in the utility.
Next, the DMS evaluates whether the OESs from various prosumers are \textit{electrically feasible} at a system level considering AC voltage and flow constraints. 
If found infeasible, the DMS communicates adjusted export setpoints (AES) to the prosumers while considering equity and sparsity in its curtailment design. 

\subsection{Receding Horizon Multi-period Optimization in HEMS}\label{subsec:math-formulation-hems}
\noindent In the ANOCA approach, HEMS optimizes energy costs (similar to \cite{yi2022fair}) by solving for an optimal battery charging and discharging behavior, given a forecast of solar PV output and electric consumption. 
Based on optimal battery charge and discharge decisions and forecasted PV values, the HEMS algorithm calculates optimal export and import set points (OES/OIS) for each time interval $\tau \in \mathcal{T} = \{t, t + \Delta t, t + 2\Delta t, \ldots, t +H\}$, where $t$ is the current time step and $H$ is the planning horizon.
HEMS achieves its goal via receding horizon multiperiod optimization in \eqref{eq:HEMS_problem} that it runs over the entire time-horizon $H$ with time interval $\Delta t$ (say, 5-15 min). 
Note while the algorithm considers look-ahead constraints based on load and PV forecasts for all time intervals $\tau \in \mathcal{T}$ into the future, it only broadcasts the optimal export setpoint (OES) for current time $t$ to the utility DMS (described in Section \ref{subsec:math-form-DMS}).
In essence, it takes a moving horizon approach and solves a recursive look-ahead multi-period optimization for each time interval.
In ANOCA, every prosumer $p \in \mathcal{P}$ runs their HEMS independently and in parallel for every time interval they wish to export power. 
Next, we mathematically define the HEMS problem.
\begin{subequations}\label{eq:HEMS_problem}
\setlength{\abovedisplayskip}{0.5em} 
\setlength{\belowdisplayskip}{0.5em} 
\setlength{\jot}{0.2em} 
\begin{alignat}{3}
    &\min_{P_{c;d}, P_{i;e}} \sum_{\tau \in T} \left( C_{i,\tau}^{p} \cdot P_{i,\tau}^{p} - C_{e,\tau}^{p} \cdot P_{e,\tau}^{p} \right)\label{eq:HEMS_objective} \\
    &\textrm{subject to:} \notag\\
    &P_{l,\tau}^{p} + P_{c,\tau}^{p} + P_{e,\tau}^{p} = P_{i,\tau}^{p} + P_{pv,\tau}^{p} + P_{d,\tau}^{p}, \; \quad \forall \tau \in \mathcal{T} \label{eq:Power_balance} \\ 
    &E_{b,\tau+1}^{p} = E_{b,\tau}^{p} \!+\! \eta_{c}^{p} P_{c,\tau}^{p} \Delta \tau \!-\! (P_{d,\tau}^{p} \Delta \tau)/\eta_{d}^{p}, \quad \forall \tau \in \mathcal{T} \label{eq:SOC_dynamics} \\
    &E_{b,0}^{p} = E_{b, \text{set}}^{p} = E_{b, t+H}^{p} \label{eq:initial_SOC} \\
    &0 \leq E_{b,\tau}^{p} \leq \overline{E}^{p}, \hspace{42mm}  \forall \tau \in \mathcal{T} \label{eq:SOC_limits} \\
    &0 \leq P_{c,\tau}^{p} \leq z_{\tau}^{p} \cdot \overline{P}^{p}, \hspace{36mm}  \forall \tau \in \mathcal{T} \label{eq:MILP_charging_limits} \\
    &0 \leq P_{d,\tau}^{p} \leq (1 - z_{\tau}^{p}) \cdot \overline{P}^{p}, \hspace{27mm}  \forall \tau \in \mathcal{T} \label{eq:MILP_discharging_limits} \\
    &z_{\tau}^{p} \in \{0,1\}, \hspace{47mm}  \forall \tau \in \mathcal{T} \label{eq:MILP_binary_variables}
\end{alignat}
\end{subequations}

\begin{figure}[t]
    \centering
    \includegraphics[width=0.8\linewidth]{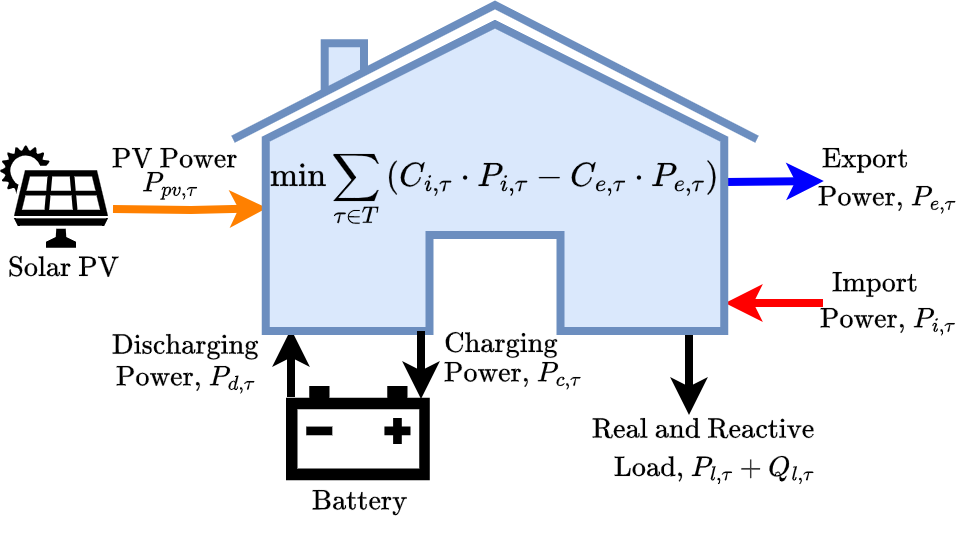}
    \caption{Schematic representation of the HEMS optimization model incorporating prosumers with solar PV and battery storage systems.}
    \label{fig:HEMS}
\end{figure}

\noindent We formulate the problem in \eqref{eq:HEMS_problem} as a mixed integer linear program (MILP). 
We include operational constraints of power balance \eqref{eq:Power_balance} and battery charge/discharge dynamics \eqref{eq:SOC_dynamics}, including state-of-charge (SOC) management for all time-intervals $t \in T$. We use this approach over NLP battery formulation because \cite{elsaadany2023battery} shows that MILP formulation outperforms NLP formulation when problems include energy-reservoir battery models.
The problem's objective is to minimize the energy cost over the time horizon $H$ for all prosumers connected to the grid with solar PV and battery system (\ref{eq:HEMS_objective}) as shown in Fig. \ref{fig:HEMS}. 
We assume the cost to import unit power $C_{i, \tau}$ for all time steps $\forall \tau \in \mathcal{T}$ is always higher than the cost to export unit power $C_{e, \tau}$.
This prevents prosumers from buying electricity from the grid and selling it at an advantage back in real-time.
The operational limits of the battery, defined in \eqref{eq:SOC_limits}, \eqref{eq:MILP_charging_limits}, and \eqref{eq:MILP_discharging_limits}, ensure the SOC remains within technical bounds.

\begin{figure}[b]
    \centering\includegraphics[width=1\linewidth]{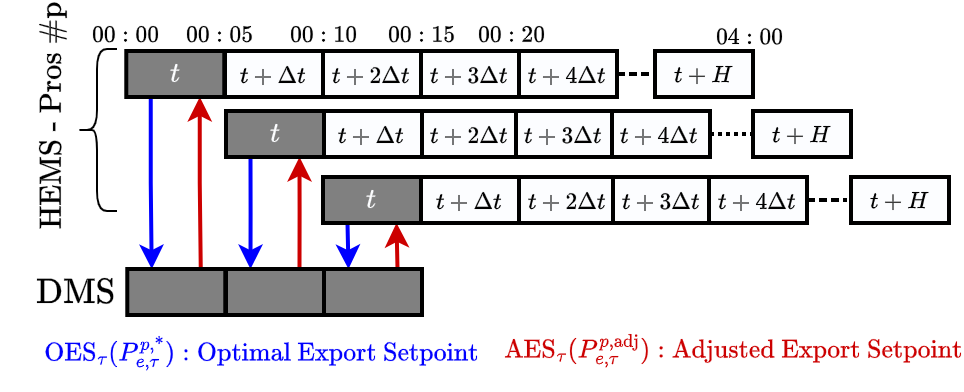}
    \caption{Interactions with HEMS receding horizon problem and DMS problem. $\tau \in T = \{t, t + \Delta t, t + 2\Delta t, \ldots, t + H\}$ represents discrete time intervals over a planning horizon $H$.}
    \label{fig:RH_algo}
\end{figure}

\begin{remark}
    \textit{HEMS calculates optimal import and export power setpoints ($P_{i,\tau}^{p,*}, P_{e,\tau}^{p,*}$) for each time step $\tau \in T$ based on the load and solar forecasts ($P_{l,\tau}^{p}, P_{pv,\tau}^{p}$) over a time horizon $H$. The forecasts are stochastic, and we minimize the risk due to underlying uncertainty by i) choosing a small time-interval $\Delta t$, usually 5/15 minutes, where the forecasts can be considered accurate and ii) only broadcasting optimal export and import setpoints for a current time step $t$ to DMS and discarding the rest, following a receding horizon look-ahead approach (see Fig. \ref{fig:RH_algo}).} 
\end{remark}

\subsection{Utility DMS-based AC-Network Aware Dispatch Certificate Optimization}\label{subsec:math-form-DMS}

\noindent In ANOCA, DMS operates at a systems level, assessing the collective impact of individual prosumer export proposals on grid stability.
Once the various prosumers request to dispatch optimal export setpoints (OES), the DMS formulates and solves an optimization to either i) allow the requested OES dispatch by prosumers (no curtailment) or ii) enforce a new adjusted export setpoint (AES) dispatch value (requesting an optimal curtailment by HEMS).
DMS problem is network-aware and operates on a full AC three-phase network model and considers operational voltage and flow bounds (similar to concepts in \cite{foster2022three}).
It uses available control actions (such as transformer taps) to manage grid constraints, and curtailment via adjusted export setpoint is used only as a last resort.
Importantly, because we consider the full AC model of the distribution grids, we incur no drawbacks on earlier works \cite{liu2022using}\cite{yi2022fair}.
At a high level, the DMS problem minimizes the norm of curtailable power $P_{cu}$ in the network to enforce AC feasibility.
The various curtailment strategies are based on the choice of metric that quantifies the net power curtailment.
We consider two broad strategies: i) a low cardinality curtail option, where only a few large curtailments are sufficient to operate the grid reliably, and ii) an equitable curtailment strategy, where the curtailments are spread more equitably amongst all prosumers (by capacity). Next, we describe the mathematical formulation of the DMS problem.

\vspace{-1.5em}
\begin{subequations}\label{eq:DMS_problem}
\setlength{\abovedisplayskip}{0.5em} 
\setlength{\belowdisplayskip}{0.5em} 
\setlength{\jot}{0.2em} 
\begin{alignat}{3}
    \min_{x \in X} \sum_{n \in \mathcal{N}_{pv}} \sum_{\phi \in \Phi} w_{n,\phi} \cdot \lVert P_{cu,n,\phi} \rVert_y
    \label{DMS_objective}
\end{alignat}
\textrm{subject to:}\\
\textrm{\underline{KCL-based network constraints}}
\begin{alignat}{3}\label{eq:KCL_start}
    &\sum_{m\in\mathcal{N}}\sum_{\phi'\in\Phi}\!(V_{nm,\phi'}^RG_{nm,\phi,\phi'}\!-\!V_{nm,\phi'}^IB_{nm,\phi,\phi'})\!+I_{l,n,\phi}^R\!=\!0, \nonumber\\
    &\hfill\hspace{5cm}\forall n\in\mathcal{N},\forall \phi\in\Phi\\
    &\sum_{m\in\mathcal{N}}\sum_{\phi'\in\Phi}\!(V_{nm,\phi'}^RB_{nm,\phi,\phi'}\!+\!V_{nm,\phi'}^IG_{nm,\phi,\phi'})\!+I_{l,n,\phi}^I\!=\!0,\nonumber\\
    &\hfill\hspace{5cm}\forall n\in\mathcal{N},\forall \phi\in\Phi\\
    &I^{R}_{l,n,\phi} = \frac{P_{n,\phi} V^{R}_{n,\phi} + Q_{n,\phi} V^{I}_{n,\phi}}{(V^{R}_{n,\phi})^2 + (V^{I}_{n,\phi})^2}, \quad \forall n \in \mathcal{N}_l, \forall \phi \in \Phi \\
    &I^{I}_{l,n,\phi} = \frac{P_{n,\phi} V^{I}_{n,\phi} - Q_{n,\phi} V^{R}_{n,\phi}}{(V^{R}_{n,\phi})^2 + (V^{I}_{n,\phi})^2}, \quad \forall n \in \mathcal{N}_l, \forall \phi \in \Phi \label{eq:KCL_end}
\end{alignat}
\noindent \textrm{\underline{Nodal power constraints including curtailment}}
\begin{alignat}{3}
    &P_{n,\phi} = P_{i,n,\phi}^{*} - P_{e,n,\phi}^{*} + P_{cu,n,\phi}, \forall n \in \mathcal{N}_{\text{pv}}, \forall \phi \in \Phi  \label{eq:Curtailment_start} \\
    &P_{n,\phi} = P_{l,n,\phi}, \hspace{21mm} \forall n \in \mathcal{N}_{\text{l}} \setminus \mathcal{N}_{\text{pv}}, \forall \phi \in \Phi \label{eq:Load} \\
    &Q_{n,\phi} = Q_{l,n,\phi}, \hspace{29mm} \forall n \in \mathcal{N}_{\text{l}}, \forall \phi \in \Phi \label{eq:Q_Load} \\
    &0 \leq P_{cu,n,\phi} \leq P_{e,n,\phi}^{*}, \hspace{18mm} \forall n \in \mathcal{N}_{\text{pv}}, \forall \phi \in \Phi \label{eq:Curtailment_end}
\end{alignat}
\noindent \underline{Flow and nodal voltage bounds}
\begin{alignat}{3}
    &(I^{R}_{e, fl,\phi})^2 + (I^{I}_{e, fl,\phi})^2 \leq (\overline{I}_{e, fl,\phi})^2, \hspace{10mm} \forall e \in \text{lines} \label{eq:current_bounds} \\
    & \underline{t}_{k} \leq t_k \leq \overline{t}_{k}, \hspace{39mm} \forall k \in \text{xfmrs} \label{eq:tap_bounds} \\
    & (P_{k,fl})^2 + (Q_{k,fl})^2 \leq (\overline{S}_{k,fl})^2, \hspace{13mm} \forall k \in \text{xfmrs} \label{eq:power_bounds} \\
    &\underline{V}_{n,\phi}^2\!\leq\!(V_{n,\phi}^R)^2\!+\!(V_{n,\phi}^I)^2\!\leq\!\overline{V}_{n,\phi}^2,\hspace{6mm} \forall n\!\in\!\mathcal{N},\forall\phi\!\in\!\Phi \label{eq:voltage_bounds_lines}
\end{alignat}
\end{subequations}
\noindent Finally, as a post-processing step, the DMS calculates and communicates the adjusted power setpoint (AES) to each  prosumer for the current time step $t$:
\vspace{-0.5em}
\begin{alignat}{3}
    P_{e,t}^{p,adj}\!=\!P^{p,*}_{e,t}\!-\!P_{cu,p,t},\;p\!=\!\{n,\phi\},\forall n\!\in\!\mathcal{N}_{pv},\forall\phi\!\in\!\Phi
\end{alignat}
\vspace{-0.25em}
\noindent In \eqref{eq:DMS_problem}, $X$ is the set of decision and dependent variables, including nodal voltages $(V^R, V^I)$, curtailment power $(P_{cu})$, line $(I_{e,fl})$ and transformer flows $(S_{k,fl})$, transformer taps ($t_k$) and others. $y$ in the objective represents the choice of norms (see subsection \ref{subsec:Norms}).
The model minimizes the ``curtailment'' in the problem objective (\ref{DMS_objective}). 
The introduction of weights \(w_{n,\phi}\) in the objective function (\ref{DMS_objective}) serves to prioritize the curtailment across different nodes and phases, ensuring that the optimization process not only seeks to minimize overall curtailment but does so in a manner that aligns with the grid's operational priorities.
Curtailments are nonnegative, as indicated in constraint (\ref{eq:Curtailment_end}), and are bounded by the optimal export setpoint proposed by HEMS. 

This problem is subject to the KCL constraints for both the real and imaginary parts (\ref{eq:KCL_start}-\ref{eq:KCL_end}).
\eqref{eq:Curtailment_start}-\eqref{eq:Q_Load} model the power balance constraints at prosumer and non-prosumer nodes, respectively.
Flow limits for lines and transformers are included in (\ref{eq:current_bounds}) and (\ref{eq:power_bounds}), respectively. 
In this paper, the model assumes transformer tap ratios as continuous variables within a specific minimum and maximum value to regulate voltage, as outlined in (\ref{eq:tap_bounds}). 
However, the authors note that transformer taps are generally discrete settings in real world. 
Methods presented in \cite{mcnamara2022two} can be readily adapted to this algorithm to accommodate and ensure discrete tap settings, bridging the gap between the simplified model and real-world operational constraints. 
The voltage levels at the load buses and transformers are bounded in (\ref{eq:voltage_bounds_lines}). 
\vspace{-0.4em}
\noindent \subsubsection*{\textbf{Curtailment Strategies}}\label{subsec:Norms}
\noindent DMS minimization allows for various curtailment strategies by selecting different norms in the objective. 
This section explores three norm-based curtailment strategies: \Lone, \Ltwo, and \Linf.
The \Lone-norm strategy aims to minimize the absolute sum of curtailments across all prosumers, resulting in curtailment for a sparse number of prosumers.
The \Ltwo-norm method seeks to minimize the sum of the square of curtailments, resulting in a fairer distribution of power curtailments (i.e., net curtailment is shared amongst prosumers). 
The \Linf-norm strategy minimizes the maximum curtailment any prosumer faces, ensuring that no individual prosumer is disproportionately affected.
The three strategies result in different curtailments amongst prosumers and present different computational complexity.
Next, we discuss the mathematical representation of different curtailment strategies in ANOCA DMS algorithm.

\noindent \textbf{\Lone{} formulation:} results in non-zero curtailments at a sparse set of nodes. We formulate a differentiable \Lone-norm problem in \eqref{eq:l1_formulation} because the naive formulation is non-differentiable and cannot be directly used with interior point solvers.
\vspace{-0.2em}
\begin{subequations} \label{eq:l1_formulation}
\setlength{\abovedisplayskip}{0.5em} 
\setlength{\belowdisplayskip}{0.5em} 
\setlength{\jot}{0.2em} 
\begin{alignat}{3}
    \min_{x \in X} \sum_{n \in \mathcal{N}_{\text{pv}}} \sum_{\phi \in \Phi} w_{n,\phi}\cdot\left(P^{+}_{cu,n,\phi}\!+\!P^{-}_{cu,n,\phi}\right)\label{DMS_objective_L1}
\end{alignat}
\textrm{subject to:}
\begin{alignat}{3}
    &~\eqref{eq:KCL_start}\!-\!\eqref{eq:KCL_end}\quad\&\quad~\eqref{eq:current_bounds}\!-\!\eqref{eq:voltage_bounds_lines}\\
    &P_{n,\phi} = P_{i,n,\phi} - P_{e,n,\phi} + P^{+}_{cu,n,\phi} - P^{-}_{cu,n,\phi}, \nonumber \\
    &\hspace{5cm} \forall n \in \mathcal{N}_{\text{pv}}, \forall \phi \in \Phi \\
    &P_{n,\phi}=P_{l,n,\phi},\hspace{24mm} \forall n\!\in\!\mathcal{N}_{\text{l}}\setminus\mathcal{N}_{\text{pv}},\forall\phi\!\in\!\Phi\label{eq:Load_L1}\\
    &P^{+/-}_{cu,n,\phi}\geq0, \hspace{32mm} \forall n \in \mathcal{N}_{\text{pv}}, \forall \phi \in \Phi \label{eq:non_negative_L1}\\
    &0\leq P^{+}_{cu,n,\phi}\!-\!P^{-}_{cu,n,\phi}\leq P_{e,n,\phi},\hspace{7mm}\forall n\!\in\!\mathcal{N}_{\text{pv}},\forall\phi\!\in\!\Phi\label{eq:Curtailment_end_L1}
\end{alignat}
\end{subequations}
\noindent Note that the formulation can be further simplified by setting $P^{-}_{cu,n,\phi} \rightarrow 0$ as negative curtailment is not physically viable.

\vspace{0.3em}
\noindent \textbf{\Ltwo{} formulation:} aims to distribute non-zero curtailments \textit{more} evenly by penalizing the square of curtailments; large curtailments are disincentivized.
\begin{subequations}
\setlength{\abovedisplayskip}{0.5em} 
\setlength{\belowdisplayskip}{0.5em} 
\setlength{\jot}{0.2em} 
\begin{alignat}{3}
    \min_{x \in X} \sum_{n \in \mathcal{N}_{\text{pv}}} \sum_{\phi \in \Phi} w_{n,\phi}\cdot(P_{cu,n,\phi})^2\label{DMS_objective_L2}
\end{alignat}
\textrm{subject to:}
\begin{alignat}{3}
    & ~\eqref{eq:KCL_start} - \eqref{eq:KCL_end} \quad \& \quad ~\eqref{eq:current_bounds}-\eqref{eq:voltage_bounds_lines}
\end{alignat}
\begin{alignat}{3}
    &P_{n,\phi}=P_{i,n,\phi}\!-\!P_{e,n,\phi}\!+\!P_{cu,n,\phi},\hspace{4mm}\forall n\!\in\!\mathcal{N}_{\text{pv}},\forall\phi\!\in\!\Phi\label{eq:L2_Curtailment_start}\\
    &P_{n,\phi}=P_{l,n,\phi},\hspace{22.5mm} \forall n\!\in\!\mathcal{N}_{\text{l}}\setminus\mathcal{N}_{\text{pv}},\forall\phi\!\in\!\Phi\label{eq:L2_Load}\\
    &0\leq P_{cu,n,\phi}\leq P_{e,n,\phi},\hspace{19mm} \forall n\!\in\!\mathcal{N}_{\text{pv}},\forall\phi\!\in\!\Phi\label{eq:L2_Curtailment_end}
\end{alignat}
\end{subequations}
\vspace{-0.2em}
\noindent \textbf{\Linf{} formulation:} aims to be the most equitable curtailment strategy because it minimizes the maximum curtailment candidate. Note, like \Lone-norm; we formulate the differentiable form of the problem in \eqref{eq:linf_formulation}.
\begin{subequations} \label{eq:linf_formulation}
\setlength{\abovedisplayskip}{0.5em} 
\setlength{\belowdisplayskip}{0.5em} 
\setlength{\jot}{0.2em} 
\begin{alignat}{3}
    \min_{{x \in X}}\quad&\overline{P}_{cu}\label{DEMS_objective_inf_norm}
\end{alignat}
\textrm{subject to:}
\begin{alignat}{3}
    &~\eqref{eq:KCL_start}\!-\!\eqref{eq:KCL_end}\quad\&\quad~\eqref{eq:current_bounds}\!-\!\eqref{eq:voltage_bounds_lines}\\
    &P_{n,\phi}=P_{i,n,\phi}\!-\!P_{e,n,\phi}\!+\!P_{cu,n,\phi}, \quad \forall n\!\in\!\mathcal{N}_{\text{pv}},\forall\phi\!\in\!\Phi\label{eq:Linf_Curtailment_start}\\
    &P_{n,\phi}=P_{l,n,\phi},\hspace{22mm} \forall n\!\in\!\mathcal{N}_{\text{l}}\setminus\mathcal{N}_{\text{pv}},\forall\phi\!\in\!\Phi\label{eq:Linf_Load}\\
    &-\overline{P}_{cu}\leq w_{n,\phi}\cdot P_{cu,n,\phi}\leq\overline{P}_{cu}, \hspace{4mm} \forall n\!\in\!\mathcal{N}_{\text{pv}},\forall\phi\!\in\!\Phi\label{eq:Linf_curtailment}\\
    &P_{cu,n,\phi}\leq P_{e,n,\phi},\hspace{25mm} \forall n\!\in\!\mathcal{N}_{\text{pv}},\forall\phi\!\in\!\Phi\label{eq:Linf_Curtailment_end}
\end{alignat}
\end{subequations}
\noindent $\overline{P}_{cu}$ represents the maximum weighted curtailment across all nodes $n \in \mathcal{N}_{\text{pv}}$ and phase $\phi$. Note that \eqref{eq:Linf_curtailment} is further reduced in implementation because $P_{cu,n,\phi} \geq 0$. 


The choice among \Lone, \Ltwo, and \Linf-norm formulations depends on the specific goals of the distribution system operator, such as minimizing total curtailment, ensuring equitable distribution, or having few flexible commercial customers do all the curtailment. 
\vspace{-0.5em}

\subsection{ANOCA Algorithm}

\noindent The two-stage optimization ANOCA is described in Algorithm \ref{alg:Efficient_HEMS_DMS_Curtailment} with HEMS at the prosumer premises and DMS at the utility. 
The DMS aggregates the optimal export setpoint (OES) requests from the HEMS for the current time step $t$. 
It ascertains whether these exports can be accommodated without surpassing grid limits.
Based on the optimization result, it issues either a go-ahead signal with no curtailment or an adjusted export setpoint (AES) signal with optimal curtailment enforcement.
The entire adjustment and communication process occurs within a predefined time interval, $\Delta t$, enabling the real-time capability of the system to respond to grid conditions and maintain operational stability. 




\begin{algorithm}[htpb]
\caption{ANOCA}
\label{alg:Efficient_HEMS_DMS_Curtailment}
\begin{algorithmic}[1] 

\State \textbf{Initialize:} $t =0$, $\Delta t = 5$ min, planning horizon $H$ = 4 hours
\For{each time $t \rightarrow$ until in operation}
\State \textit{\textbf{Initiate} HEMS Receding Horizon Multi-Period \State Optimization}
    \For{each prosumer $p \in P$ in parallel:}
    \Indent
        \State \textbf{Input:} Battery  parameters, forecasted load \State $P_{l,\tau}^n$ and solar $P_{pv,\tau}^n$ output  for time intervals \State $\tau \in T = \{t, t + \Delta t, t + 2\Delta t, \ldots, t +H\}$

        \State \textbf{Run:} Receding-horizon multi-period \State optimization problem in \eqref{eq:HEMS_problem} $\rightarrow$ $P^{p,*}_i, P^{p,*}_e$

        \State \textbf{Broadcast:} Optimal export setpoint \textit{OES} \State $(P^{p,*}_{e,t})$ and optimal import setpoint \textit{OIS} 
        \State $(P^{p,*}_{i,t})$  of  $p \in \mathcal{P}$ to DMS
    
    \EndIndent
    \EndFor
\State \textit{\textbf{Initiate} DMS AC Network-aware Dispatch \State  Certificate Optimization}
\State \textbf{Input:} Broadcasted \textit{OES} and \textit{OIS} setpoints from all  \State  prosumers $(P^*_{e,t}, P^*_{i,t})$, loads for all nodes,  grid model \State for current time $t$
\State \textbf{Select} a curtailment strategy $\leftarrow$ (\Lone, \Ltwo, \Linf)
\State \textbf{Run} DMS-based AC optimization in \eqref{eq:DMS_problem}
\If{output $P_{cu,p,t} = 0$ for all prosumers $p \in P$}
    \State No curtailment needed; communicate adjusted  \State export  setpoints, \textit{AES} $=$ \textit{OES}, for all prosumers
\Else
    \State Calculate adjusted export  setpoints (\textit{AES}) for  \State prosumers: $P_{e,t}^{p,adj} = P^{p,*}_{e,t} - P_{cu,p,t}$
\EndIf
\State \textbf{Communicate} \textit{AES}: $P_{e,t}^{p,adj}$  to all prosumers $\forall p \in P$
\State \textbf{Update} HEMS setpoints for time $t$
\For{each prosumer $p \in P$:}
    \State \textbf{Update} HEMS export setpoint to $P_{e,t}^{p,adj}$
\EndFor
\EndFor
\end{algorithmic}
\end{algorithm}

%
%
%


\section{Experimental Setup} \label{sectionIV}
\noindent 
Next, we describe the experimental setup to study and validate the efficacy of the ANOCA framework:
\subsection{HEMS Experiment Data}
\noindent For HEMS experiments, we obtain load consumption and solar output data from a realistic feeder in New York \cite{almassalkhi2020hierarchical}. 
We generate our forecasts by mimicking the load and solar pattern in this data and adding Gaussian noise.
We assume price tariffs based on data from the International Energy Agency (IEA) \cite{IEA2022Electricity}, reflecting New York Independent System Operator (NYISO) standards. 
We employ Tesla Powerwall 2 as the battery system parameters in our studies \cite{TeslaPowerwall2Datasheet2023}. 
The simulations employed a 5-minute timestep granularity, with the battery's start and end State of Charge (SOC) setpoints at 50\% of its maximum capacity. 
\subsection{DMS Experiment Data}\label{cases}
\noindent We study two network cases (across 4 scenarios) to evaluate the performance of the proposed DMS algorithm. Each case has three categories of prosumers, and the net PV output (as a $\%$ of base load) for various scenarios is in Table \ref{table:scenario_analysis}.

\begin{itemize}
    \item \textbf{Case 1:} Standard three-phase IEEE-4 bus network \cite{IEEE2006FourNode}
    \item \textbf{Case 2:} A synthetic urban-meshed network, with 1420 3$\phi$ nodes (4260 
    1$\phi$ nodes) and 624 load buses \cite{schneider2014ieee}.
\end{itemize}
\begin{table}[htbp] 
    \caption{Scenario Description: PV Export ($\%$) of Base Load}
    \centering
    \setlength{\tabcolsep}{2.5pt} 
    \renewcommand{\arraystretch}{0.5} 
    \begin{tabular}{@{}c|ccc@{}}
        \toprule
        \textbf{Scenario ID} & \textbf{Category A ($\%$)} & \textbf{Category B ($\%$)} & \textbf{Category C ($\%$)} \\
        \midrule
        1 & 150 & 0 & 30 \\
        2 & 30 & 150 & 0 \\
        3 & 30 & 0 & 150 \\
        4 & 0 & 150 & 30 \\
        \bottomrule
    \end{tabular}
    \label{table:scenario_analysis}
\end{table}

\subsection{Simulation Environment and Hardware}
\noindent The experimental simulations were done on a PC with a 2.7 GHz Intel Core i7 processor and 16 GB RAM. IPOPT (version 3.14.10) \cite{wachter2006implementation} was used for DMS optimization tasks. Gurobi (Version 11.0.0) \cite{gurobi2021gurobi} was used for HEMS optimization. Our data used are publicly available at \url{https://github.com/emmanuelbadmus/ANOCA}

\section{Experiments}
\noindent This section documents the outcomes of the experiments with two-stage ANOCA optimization framework. 
We show that the ANOCA framework can accommodate varying power exports from DERs while maintaining grid stability and ensuring adherence to voltage safety limits. We observe widely varying curtailment patterns dependent on the choice of objective function.

\subsection{HEMS Analysis in ANOCA}
\noindent Here, we evaluate the HEMS problem in \eqref{eq:HEMS_problem} within the ANOCA framework.
We choose a look-ahead time horizon $(H)$ of 8 hours, with an individual dispatch time interval of 5 minutes. 
Therefore, the HEMS multi-period problem runs over a total of 96 intervals. 
The total number of prosumers in \textbf{Cases 1} and \textbf{2} is two (2) and four hundred and sixteen (416), respectively. 
We used Gurobi to solve the HEMS problem. 
The average solution time for HEMS optimizations was 0.107 seconds over 12 scenarios, confirming that the ANOCA framework is not time-constrained. 
The duality gap for each scenario was 0.0\%, validating that we converge to a global minimum for all multi-period problems. 
We communicate the optimal export setpoints from the HEMS to DMS analysis for time step $t$.


\subsection{DMS Analysis in ANOCA}

\subsubsection{Scenario with No Exports}
\noindent In the first experiment, we examine \textbf{Case 1} when none of the prosumers are exporting power. 
This enables us to validate the ANOCA DMS algorithm against the power flow solution from GridLab-D \cite{chassin2008gridlab} as the DMS optimization problem in \eqref{eq:DMS_problem} is equivalent to the non-binding power flow run when prosumers are not exporting any power. 
We compare the load bus voltages from ANOCA against GridLAB-D simulation results and document them in Table \ref{table:voltage_comparison_1}. 
The deviations in voltage levels are within numerical tolerance, validating the basic functionality of the DMS algorithm in ANOCA.

\begin{table}[htbp] 
    \caption{Comparison of voltage levels using ANOCA and GridLAB-D}
    \centering
    \setlength{\tabcolsep}{5pt} 
    \renewcommand{\arraystretch}{0.5} 
    \begin{tabular}{@{}c|ccc@{}}
        \toprule
        \textbf{Load ID} & \textbf{Gridlab-D, (V)} & \textbf{ANOCA, (V)} & \textbf{Difference (\%)} \\
        \midrule
        0 & 1917.76 & 1915.83 & -0.101\% \\
        1 & 2061.26 & 2069.40 & +0.394\% \\
        2 & 1980.76 & 1966.38 & -0.729\% \\
        \bottomrule
    \end{tabular}
    \label{table:voltage_comparison_1}
\end{table}

\subsubsection{Incremental Power Export and Voltage Stability}
\noindent Next, we conduct an experiment on \textbf{Case 1},  where ANOCA enforces curtailments to maintain voltage limits $[0.95-1.05]$. To achieve this, we allow prosumers to export power under scenario 3. We apply \Lone{} curtailment strategy, and we document the load bus voltages $(|V|)$ with and without ANOCA and the amount of 
power curtailed $(P_{cu})$ in Table \ref{table:voltage_comparison_2}. The results show that ANOCA can apply curtailments optimally to maintain technical bounds.

\begin{table}[htbp]
    \caption{Comparison of Voltage with/without ANOCA}
    \centering
    \footnotesize 
    \setlength{\tabcolsep}{10pt} 
    \renewcommand{\arraystretch}{0.5} 
     \begin{tabular}{@{}c|ccc@{}}
        \toprule
        \textbf{Load ID} & \textbf{W/O ANOCA} & \multicolumn{2}{c}{\textbf{W/ ANOCA}}  \\
        & $|V|$ [p.u] & $|V|$ [p.u] & $P_{cu}$ [\%Load] \\
        \midrule
        0 & 1.22 & 1.05 & 16.28 \\
        1 & 1.15 & 1.01 & 134.05 \\
        2 & 0.79 & 0.98 & 0.00 \\
        \bottomrule
    \end{tabular}
    \label{table:voltage_comparison_2}
\end{table}


\begin{figure*}[b]
    \centering
    \includegraphics[width=1\linewidth]{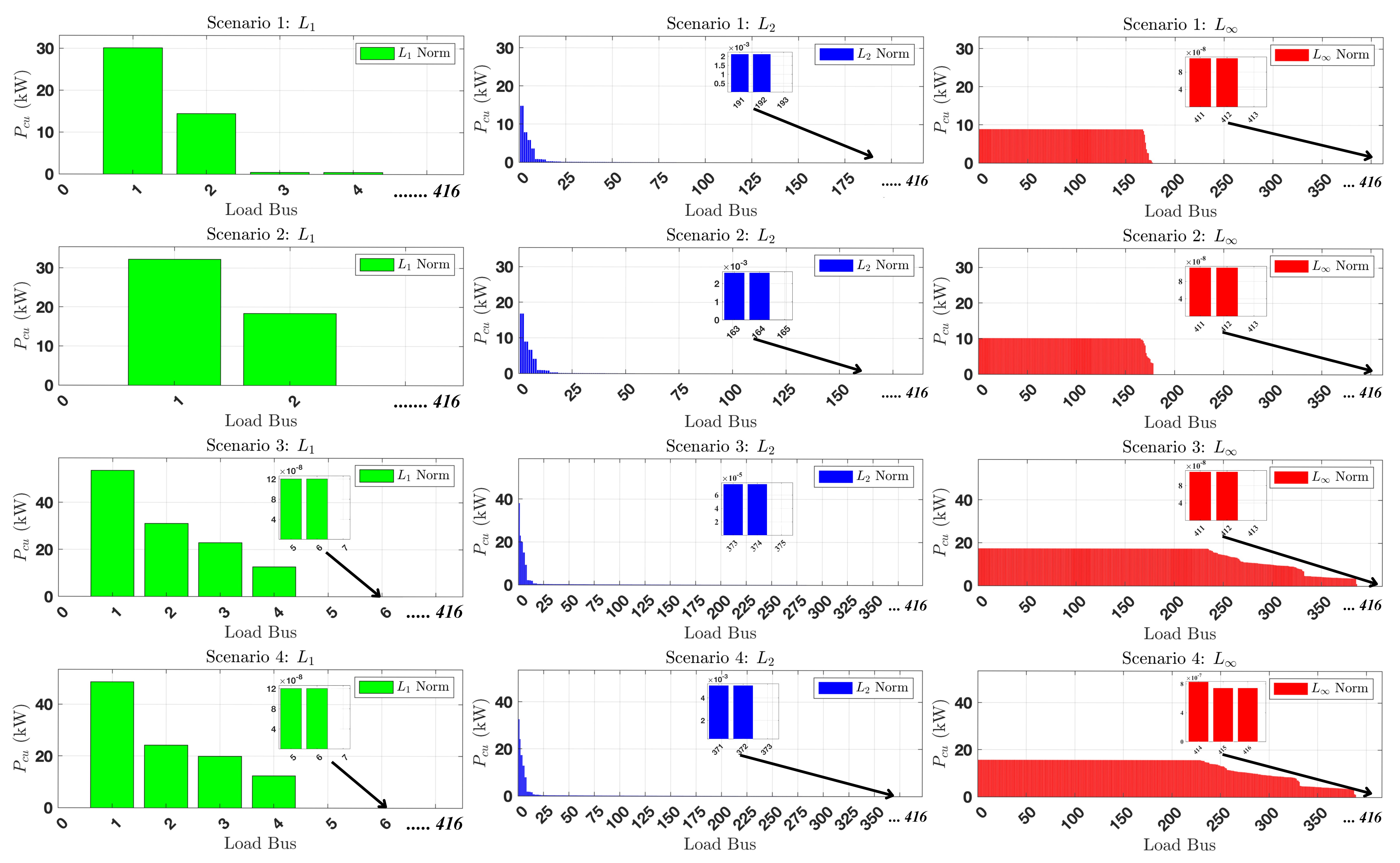}
    \caption{Comparative analysis of curtailment strategies: \Lone, \Ltwo, and \Linf-norms across four scenarios for \textbf{Case 2}: The case has 416 prosumers capable of exporting power. The analysis across various scenarios shows the number of prosumers that curtail power and the corresponding magnitude for all strategies. Notably, the \Lone-norm strategy curtails far fewer prosumer but results in a significantly higher power curtailment per prosumer than the \Ltwo{} and \Linf-norm strategies.}    
    \label{fig:Curtailment }
\end{figure*}

\subsubsection{Scaling ANOCA Results}
\noindent Next, we study the performance of ANOCA on a large, realistic-sized meshed network - \textbf{Case 2} with 1420 three-phase nodes and 4260 individual phase nodes. We apply the ANOCA framework to enforce optimal curtailments (when necessary) to maintain voltage and flow bounds. 
We explore \Lone, \Ltwo, and \Linf-norm-based curtailment strategies. 
The runtime and average iteration count results in Table \ref{table:curtailment_strategies} demonstrate the ANOCA framework's robustness with an average DMS analysis solution time on this large network of 133.83 seconds (much faster for \Lone and \Ltwo).

\subsubsection{Comparitive Assessment of Various Curtailment Strategies} 
\noindent  Now, we evaluate the effectiveness of \Lone, \Ltwo, and \Linf-norm curtailment strategies in ensuring grid reliability while optimizing PV outputs. 
We document the results in Table \ref{table:curtailment_strategies} for 4 \textbf{Case 2} scenarios.
Across all scenarios, \Lone{} curtailment strategy has the most sparse curtailment. 
On average, 2 to 6 prosumers alone can curtail enough power to maintain grid reliability.
Furthermore, the net volume of power curtailment is also the lowest for \Lone{} curtailment strategy, and it has the fastest runtime (on average, 7.26 sec).
\Linf{} curtailment strategy is the most equitable, distributing curtailment amongst almost all 416 prosumers. However, the net volume of curtailed power is the highest for \Linf{} strategy (average curtailment of 13.43$\%$ vs. 0.56$\%$ for \Ltwo{} and 0.32$\%$ for \Lone{}).
The runtime for \Linf{} is relatively more than others (on average, 15.68 seconds).  \Ltwo{} curtailment strategy strikes a balance between the strengths and drawbacks of \Lone{} and \Linf{}; it is more equitable than \Lone{} and with far lower net curtailment of \Linf{} and comparable runtime to \Lone{} (on average, 12.86 sec).

Fig. \ref{fig:Curtailment } reinforces the insights discussed in Table  \ref{table:curtailment_strategies}, illustrating the quantitative impacts under each curtailment strategy. Fig. \ref{fig:Curtailment } shows that while \Lone{} has the least number of prosumers curtailing export ($<$10), the maximum curtailment far exceeded \Linf{} and \Ltwo{} (around $\sim$ 30-53 kW). On the other hand, while \Linf{} reduced the maximum curtailment for any given prosumer (around $\sim$ 10 kW), the number of prosumers curtailing is far higher ($>$400), including net curtailment. 


\begin{table}[htpb]
\centering
\setlength{\tabcolsep}{10pt}
\renewcommand{\arraystretch}{0.5}
\caption{Comparative Analysis of ANOCA Curtailment Strategies}
\label{table:curtailment_strategies}
\begin{tabular}{@{}l|ccc@{}}
\toprule
\textbf{Metric} & \textbf{\Linf} & \textbf{\Ltwo} & \textbf{\Lone} \\
\midrule
$\#$ Pros., Cons., 3-$\Phi$ nodes (All) & \multicolumn{3}{c}{416, 624, 1420} \\
\midrule
\multicolumn{4}{c}{\textbf{Scenario 1}} \\
\midrule
Total Power (MW) & \multicolumn{3}{c}{25.36} \\
Net. Curtailed Power (MW) & 1.52 & 0.08 & 0.04 \\
Maximum Curtailed power (KW) & 8.89 & 14.78 & 30.10 \\
$\#$ Load Buses Adjusted & 412 & 192 & 4 \\
\midrule
\multicolumn{4}{c}{\textbf{Scenario 2}} \\
\midrule
Total Power (MW) & \multicolumn{3}{c}{25.54} \\
Net. Curtailed Power (MW) & 1.75 & 0.08 & 0.04 \\
Maximum Curtailed power (KW) & 10.12 & 16.80 & 32.18 \\
$\#$ Load Buses Adjusted & 412 & 164 & 2 \\
\midrule
\multicolumn{4}{c}{\textbf{Scenario 3}} \\
\midrule
Total Power (MW) & \multicolumn{3}{c}{25.05} \\
Net. Curtailed Power (MW) & 5.40 & 0.21 & 0.12 \\
Maximum Curtailed power (KW) & 17.38 & 38.10 & 53.57 \\
$\#$ Load Buses Adjusted & 412 & 374 & 6 \\
\midrule
\multicolumn{4}{c}{\textbf{Scenario 4}} \\
\midrule
Total Power (MW) & \multicolumn{3}{c}{25.46} \\
Net. Curtailed Power (MW) & 4.92 & 0.19 & 0.11 \\
Maximum Curtailed power (KW) & 15.70 & 32.59 & 48.66 \\
$\#$ Load Buses Adjusted & 416 & 372 & 6 \\
\midrule
\multicolumn{4}{c}{\textbf{Average across all scenarios}} \\
\midrule
\% Avg. Curtailment & 13.43\% & 0.56\% & 0.32\% \\
\% Avg. Load Buses Adjusted & 99.28\% & 66.23\% & 1.08\% \\
\midrule
Avg. time - 4 scenerios (sec) & 15.68 & 12.86 & 7.28 \\
Avg. Iter. \# - 4 scenerios & 268 & 177 & 120 \\
\bottomrule
\end{tabular}
\end{table}



\section{Conclusions}
\noindent We introduce a two-stage optimization framework, ANOCA, to address gaps in the current dynamic hosting capacity paradigm. 
We use the \IV-based AC model of the distribution grid and, therefore, ensure that curtailments based on the proposed DHC do not violate network constraints in practical scenarios.
We also maximize network utilization by accurately capturing the boundary conditions of the distribution grid.
We conclude the following regarding the ANOCA's curtailment strategies:
\begin{enumerate}
    \item \Lone-norm curtailment strategy results in the most sparse curtailment. The strategy is most applicable when the utility has flexible customers (e.g., large computing facility).
    \item \Linf-norm curtailment strategy results in the most equitable curtailment; however, the volume of net curtailment is significantly higher than \Lone and \Ltwo.
    \item \Ltwo-norm curtailment norm strategy strikes a balance with \textit{somewhat} equitable curtailment with low net curtailment.
\end{enumerate}
We show that overall ANOCA framework relies on the same communication framework as prior work and is fast, such that both HEMS + DMS simulations solve on average in less than 20 seconds for all.



\section{Aknowledgement}
\noindent We would also like to extend our gratitude to Muhammad Hamza Ali for his assistance with the development of the code used in this research.

\bibliographystyle{ieeetr}
\bibliography{refrences}

\begin{thebibliography}{10}

\bibitem{bollen2005power}
M.~Bollen and M.~H{\"a}ger, ``Power quality: interactions between distributed energy resources, the grid, and other customers,'' {\em Leonardo Energy}, 2005.

\bibitem{capitanescu2014assessing}
F.~Capitanescu, L.~F. Ochoa, H.~Margossian, and N.~D. Hatziargyriou, ``Assessing the potential of network reconfiguration to improve distributed generation hosting capacity in active distribution systems,'' {\em IEEE Transactions on Power Systems}, vol.~30, no.~1, pp.~346--356, 2014.

\bibitem{liu2022using}
M.~Z. Liu, L.~F. Ochoa, P.~K. Wong, and J.~Theunissen, ``Using opf-based operating envelopes to facilitate residential der services,'' {\em IEEE Transactions on Smart Grid}, vol.~13, no.~6, pp.~4494--4504, 2022.

\bibitem{yi2022fair}
Y.~Yi and G.~Verbi{\v{c}}, ``Fair operating envelopes under uncertainty using chance constrained optimal power flow,'' {\em Electric Power Systems Research}, vol.~213, p.~108465, 2022.

\bibitem{astero2018improving}
P.~Astero and L.~S{\"o}der, ``Improving pv dynamic hosting capacity using adaptive controller for statcoms,'' {\em IEEE Transactions on Energy Conversion}, vol.~34, no.~1, pp.~415--425, 2018.

\bibitem{Liu_2023}
B.~Liu and J.~H. Braslavsky, ``Robust dynamic operating envelopes for der integration in unbalanced distribution networks,'' {\em IEEE Transactions on Power Systems}, p.~1–15, 2023.

\bibitem{gebbran2022multiperiod}
D.~Gebbran, S.~Mhanna, A.~C. Chapman, and G.~Verbi{\v{c}}, ``Multiperiod der coordination using admm-based three-block distributed ac optimal power flow considering inverter volt-var control,'' {\em IEEE Transactions on Smart Grid}, 2022.

\bibitem{mahmoodi2021hosting}
M.~Mahmoodi and L.~Blackhall, ``Der hosting capacity envelope in unbalanced distribution systems,'' in {\em 2021 IEEE PES Innovative Smart Grid Technologies Europe (ISGT Europe)}, pp.~1--6, IEEE, 2021.

\bibitem{moring2023inexactness}
H.~Moring and J.~L. Mathieu, ``Inexactness of second order cone relaxations for calculating operating envelopes,'' in {\em 2023 IEEE International Conference on Communications, Control, and Computing Technologies for Smart Grids (SmartGridComm)}, pp.~1--6, IEEE, 2023.

\bibitem{nazir2021grid}
N.~Nazir and M.~Almassalkhi, ``Grid-aware aggregation and realtime disaggregation of distributed energy resources in radial networks,'' {\em IEEE Transactions on Power Systems}, vol.~37, no.~3, pp.~1706--1717, 2021.

\bibitem{foster2022three}
E.~Foster, A.~Pandey, and L.~Pileggi, ``Three-phase infeasibility analysis for distribution grid studies,'' {\em Electric Power Systems Research}, vol.~212, p.~108486, 2022.

\bibitem{jereminov2016equivalent}
M.~Jereminov, D.~M. Bromberg, A.~Pandey, X.~Li, G.~Hug, and L.~Pileggi, ``An equivalent circuit formulation for three-phase power flow analysis of distribution systems,'' in {\em 2016 IEEE/PES Transmission and Distribution Conference and Exposition (T\&D)}, pp.~1--5, IEEE, 2016.

\bibitem{garcia2000three}
P.~A. Garcia, J.~L.~R. Pereira, S.~Carneiro, V.~M. Da~Costa, and N.~Martins, ``Three-phase power flow calculations using the current injection method,'' {\em IEEE Transactions on power systems}, vol.~15, no.~2, pp.~508--514, 2000.

\bibitem{jereminov2018equivalent}
M.~Jereminov, A.~Pandey, and L.~Pileggi, ``Equivalent circuit formulation for solving ac optimal power flow,'' {\em IEEE Transactions on Power Systems}, vol.~34, no.~3, pp.~2354--2365, 2018.

\bibitem{iria2022mv}
J.~Iria, P.~Scott, A.~Attarha, D.~Gordon, and E.~Franklin, ``Mv-lv network-secure bidding optimisation of an aggregator of prosumers in real-time energy and reserve markets,'' {\em Energy}, vol.~242, p.~122962, 2022.

\bibitem{petrou2021ensuring}
K.~Petrou, A.~T. Procopiou, L.~Gutierrez-Lagos, M.~Z. Liu, L.~F. Ochoa, T.~Langstaff, and J.~M. Theunissen, ``Ensuring distribution network integrity using dynamic operating limits for prosumers,'' {\em IEEE Transactions on Smart Grid}, vol.~12, no.~5, pp.~3877--3888, 2021.

\bibitem{elsaadany2023battery}
M.~Elsaadany and M.~R. Almassalkhi, ``Battery optimization for power systems: Feasibility and optimality,'' in {\em 2023 62nd IEEE Conference on Decision and Control (CDC)}, pp.~562--569, IEEE, 2023.

\bibitem{mcnamara2022two}
T.~McNamara, A.~Pandey, A.~Agarwal, and L.~Pileggi, ``Two-stage homotopy method to incorporate discrete control variables into ac-opf,'' {\em Electric Power Systems Research}, vol.~212, p.~108283, 2022.

\bibitem{almassalkhi2020hierarchical}
M.~Almassalkhi, S.~Brahma, N.~Nazir, H.~Ossareh, P.~Racherla, S.~Kundu, S.~P. Nandanoori, T.~Ramachandran, A.~Singhal, D.~Gayme, {\em et~al.}, ``Hierarchical, grid-aware, and economically optimal coordination of distributed energy resources in realistic distribution systems,'' {\em Energies}, vol.~13, no.~23, p.~6399, 2020.

\bibitem{IEA2022Electricity}
{International Energy Agency (IEA)}, ``Real-time electricity tracker.'' \url{https://www.iea.org/data-and-statistics/data-tools/real-time-electricity-tracker}, 2022.

\bibitem{TeslaPowerwall2Datasheet2023}
I.~Tesla, {\em Powerwall 2 AC Datasheet}, 2023.
\newblock Accessed: 2023-01-07.

\bibitem{IEEE2006FourNode}
{Distribution System Analysis Subcommittee}, {\em {IEEE 4 Node Test Feeder Revised}}.
\newblock {IEEE Power Engineering Society}, Knoxville, TN, 9 2006.
\newblock Revised September 19, 2006.

\bibitem{schneider2014ieee}
K.~Schneider, P.~Phanivong, and J.-S. Lacroix, ``Ieee 342-node low voltage networked test system,'' in {\em 2014 IEEE PES general meeting| conference \& exposition}, pp.~1--5, IEEE, 2014.

\bibitem{wachter2006implementation}
A.~W{\"a}chter and L.~T. Biegler, ``On the implementation of an interior-point filter line-search algorithm for large-scale nonlinear programming,'' {\em Mathematical programming}, vol.~106, pp.~25--57, 2006.

\bibitem{gurobi2021gurobi}
L.~Gurobi~Optimization, ``Gurobi optimizer reference manual (2020),'' 2021.

\bibitem{chassin2008gridlab}
D.~P. Chassin, K.~Schneider, and C.~Gerkensmeyer, ``Gridlab-d: An open-source power systems modeling and simulation environment,'' in {\em 2008 IEEE/PES Transmission and Distribution Conference and Exposition}, pp.~1--5, IEEE, 2008.

\end{thebibliography}

\end{document}